\documentclass[11pt]{article}
\usepackage{amssymb}
\usepackage{epsfig}

\newcommand{\BE}{\begin{equation}}
\newcommand{\EE}{\end{equation}}
\newcommand{\BA}{\begin{eqnarray}}
\newcommand{\EA}{\end{eqnarray}}

\begin{document}

\begin{titlepage}
\begin{center}

   {\LARGE{\bf Precision tests with a new class of dedicated ether-drift experiments}}

\vspace*{14mm} {\Large  M. Consoli and E. Costanzo}
\vspace*{4mm}\\
{
Istituto Nazionale di Fisica Nucleare, Sezione di Catania \\
Dipartimento di Fisica e Astronomia dell' Universit\`a di Catania \\
Via Santa Sofia 64, 95123 Catania, Italy \\ }
\end{center}
\begin{center}
{\bf Abstract}
\end{center}
In principle, by accepting the idea of a non-zero vacuum energy, the
physical vacuum of present particle physics might represent a
preferred reference frame. By treating this quantum vacuum as a
relativistic medium, the non-zero energy-momentum flow expected in a
moving frame should effectively behave as a small thermal gradient
and could, in principle, induce a measurable anisotropy of the speed
of light in a loosely bound system as a gas. We explore the
phenomenological implications of this scenario by considering a new
class of dedicated ether-drift experiments where arbitrary gaseous
media fill the resonating optical cavities. Our predictions cover
most experimental set up and should motivate precise experimental
tests of these fundamental issues. \vskip 20 pt {\bf PACS} :
11.30.Cp; 11.30.Qc; 03.30.+p

To appear in the European Physical Journal {\bf C}. 

\end{titlepage}

\section{Introduction}

The idea of a `condensed vacuum' is generally accepted in modern
elementary particle physics. Indeed, in many different contexts one
introduces a set of elementary quanta whose perturbative empty
vacuum state $|o\rangle$ is not the true ground state of the theory.
For instance, in the physically relevant case of the Standard Model
of electroweak interactions, the situation can be summarized by
saying that "What we experience as empty space is nothing but the
configuration of the Higgs field that has the lowest possible
energy. If we move from field jargon to particle jargon, this means
that empty space is actually filled with Higgs particles. They have
Bose condensed" \cite{thooft}. The translation from field jargon to
particle jargon can be obtained, for instance, along the lines of
Ref.\cite{mech} where the substantial equivalence between the
effective potential of quantum field theory and the energy density
of a dilute particle system was established.

For this reason, it becomes natural to ask \cite{pagano} if Bose
condensation, i.e. the spontaneous creation from the empty vacuum of
elementary spinless quanta and their macroscopic occupation of the
same quantum state, say ${\bf{k}}=0$ in some reference frame
$\Sigma$, might represent the operative construction of a "quantum
ether". This would characterize the {\it physically realized} form
of relativity and could play the role of preferred frame in a modern
Lorentzian approach.

Usually this possibility is not considered with the motivation,
perhaps, that the average properties of the condensed phase are
summarized into a single quantity that transforms as a world scalar
under the Lorentz group. For instance, in the Standard Model, the
vacuum expectation value $\langle\Phi\rangle$ of the Higgs field.

However, this does not imply that the vacuum state itself has to be
{\it Lorentz invariant}. Namely, Lorentz transformation operators
$\hat{U}'$, $\hat{U}''$,..might transform non trivially the
reference vacuum state $|\Psi^{(0)}\rangle$ (appropriate to an
observer at rest in $\Sigma$) into $| \Psi'\rangle$, $|
\Psi''\rangle$,.. (appropriate to moving observers S', S'',..) and
still, for any Lorentz-invariant operator $\hat{G}$, one would find
\BE \langle \hat{G}\rangle_{\Psi^{(0)}}=\langle
\hat{G}\rangle_{\Psi'}=\langle
\hat{G}\rangle_{\Psi''}=..\end{equation} The possibility of a
non-Lorentz-invariant vacuum state was addressed in Ref.\cite{epjc}
by considering two basically different approaches. In a first
description, by following the axiomatic approach to quantum field
theory \cite{cpt}, the vacuum is described as an eigenstate of the
energy-momentum vector. Therefore, by observing that (with the
exception of unbroken supersymmetries) there are no known
interacting theories with a vanishing vacuum energy, and using the
Poincar\'e algebra of the boost and energy-momentum operators, one
deduces that the physical vacuum cannot be a Lorentz-invariant state
and that, in any moving frame, there should be a non-zero vacuum
spatial momentum $\langle {\hat{P}_i}\rangle_{\Psi'}\neq 0$ along
the direction of motion. In this way, for a moving observer S' the
physical vacuum would look like some kind of ethereal medium for
which, in general, one can introduce a momentum density $\langle
\hat{W}_{0i}\rangle_{\Psi'}$ through the relation (i=1,2,3)
\begin{equation} \label{density} \langle {\hat{P}_i}\rangle_{\Psi'}\equiv \int
d^3x~\langle \hat{W}_{0i}\rangle_{\Psi'}  \neq 0 \end{equation} On
the other hand, in an alternative picture where one assumes the
following form of the vacuum energy-momentum tensor
\cite{zeldovich,weinberg}
\begin{equation}\label{zeld} \langle \hat{W}_{\mu\nu}\rangle_
{\Psi^{(0)}}=\rho_v ~\eta_{\mu\nu}\end{equation} ($\rho_v$ being a
space-time independent constant and $\eta_{\mu\nu}={\rm
diag}(1,-1,-1,-1)$), one is driven to completely different
conclusions. In fact, by introducing the Lorentz transformation
matrices $\Lambda^\mu_\nu$ to any moving frame S', defining $\langle
\hat{W}_{\mu\nu}\rangle_{\Psi'}$ through the relation
\begin{equation}\langle \label{cov}
\hat{W}_{\mu\nu}\rangle_{\Psi'}=\Lambda^{\sigma}{_\mu}\Lambda^{\rho}{_\nu}
~\langle\hat{W}_{\sigma\rho}\rangle_{\Psi^{(0)}}\end{equation} and
using Eq.(\ref{zeld}),
 it follows that the expectation value of $\hat{W}_{0i}$ in any
 boosted vacuum state $| \Psi'\rangle$ vanishes, just as it vanishes
in $|\Psi^{(0)}\rangle$, i.e. \begin{equation} \label{density1} \int
d^3x~ \langle \hat{W}_{0i}\rangle_{\Psi'} \equiv \langle
{\hat{P}_i}\rangle_{\Psi'}= 0 \end{equation} As discussed in
Ref.\cite{epjc}, both approaches have their own good motivations and
it is not so obvious to decide between Eq.(\ref{density}) and
Eq.(\ref{density1}) on pure theoretical grounds.

At the same time, checking the Lorentz invariance of the physical
vacuum by an explicit microscopic calculation, in the realistic case
of the Standard Model, seems to go beyond the present possibilities.
To this end, in fact, one should construct the transformed vacuum
state $|\Psi'\rangle$ by acting with the appropriate boost generator
on the reference condensed vacuum state $|\Psi^{(0)}\rangle$. Even
disposing, at least in the simplified case of spontaneous symmetry
breaking in a pure scalar theory \cite{ciancitto}, of a
non-perturbative ansatz for $|\Psi^{(0)}\rangle$, as a coherent
state expressed in terms of the creation and annihilations operators
$a^{\dagger}_{\bf p}$ and $a_{\bf p}$ of the trivial empty vacuum
state $|o\rangle$, one is faced with a serious problem: the standard
second-quantized form of the boost generators \BE \hat{M}_{0i}=i\int
{{d^3{\bf p} }\over{(2\pi)^3}} ~a^{\dagger}_{\bf p}~ \omega({\bf
p}){{\partial}\over{\partial p_i}}~ a_{\bf p} \EE is only valid for
a free-field theory. For an interacting theory, the explicit
construction of the boost generators is only known in perturbation
theory (see e.g. \cite{poincare1,poincare2} and references quoted
therein) and thus this type of approximation could hardly be trusted
in the presence of non-perturbative phenomena such as vacuum
condensation. In addition, even in perturbation theory, the
elimination of ultraviolet divergences in global operators
represents a delicate task so that only very simple theories or
low-dimensionality cases have been worked out so far. For these
reasons, deciding on the Lorentz-invariance of the condensed vacuum
of present particle physics represents a highly non-trivial problem.

Alternatively, one might argue that a satisfactory solution of the
vacuum-energy problem lies definitely beyond flat space. A non-zero
$\rho_v$, in fact, will induce a cosmological term in Einstein's
field equations and a non-vanishing space-time curvature which
anyhow dynamically breaks global Lorentz symmetry.

Nevertheless, in our opinion, in the absence of a consistent quantum
theory of gravity, physical models of the vacuum in flat space can
be useful to clarify a crucial point that, so far, remains obscure:
the huge renormalization effect that is seen when comparing the
typical vacuum-energy scales of modern particle physics with the
experimental value of the cosmological term needed in Einstein's
equations to fit the observations. For instance, the picture of the
vacuum as a superfluid explains in a natural way why there might be
no non-trivial macroscopic curvature in the equilibrium state where
any liquid is self-sustaining \cite{volovik}. In this framework, the
condensation energy of the medium plays no observable role so that
the relevant curvature effects may be orders of magnitude smaller
than those expected by solving Einstein's equations with the full
$\langle \hat{W}_{\mu\nu}\rangle_ {\Psi^{(0)}}$ as a source term. In
this perspective,  ``induced-gravity'' \cite{adler} approaches,
where gravity somehow arises from the excitations of the quantum
vacuum itself, may become natural and, to find the appropriate form
of the energy-momentum tensor in Einstein's equations, we are lead
to sharpen our understanding of the vacuum structure and of its
excitation mechanisms by starting from the physical picture of a
superfluid medium.

By following this approach, in Ref.\cite{epjc}, to explore the
possible effects of the energy-momentum flow expected in a moving
frame according to Eq.(\ref{density}), it was adopted a
phenomenological two-fluid model in which the quantum vacuum, in
addition to the main zero-entropy superfluid component, contains a
small fraction of ``normal'' fluid. This is responsible for a
non-zero $\langle \hat{W}_{0i}\rangle_{\Psi'}$ and gives rise to a
small heat flow and to an effective thermal gradient
\begin{equation} \label{gradient}
{{\partial T }\over {\partial x^i}}\equiv-{{\langle
W_{0i}\rangle_{\Psi'} }\over{\kappa_0}}
 \end{equation} Here $\kappa_0$ is an unknown parameter, introduced for
dimensional reasons, that plays the role of thermal conductivity of
the vacuum. Since its value is unknown, the effective thermal
gradient is left as an entirely free quantity whose magnitude should
be constrained by experiments.

In principle, this effective gradient could induce small convective
currents in a loosely bound system as a gaseous medium (placed in a
container at rest in the laboratory frame) and produce a slight
anisotropy of the speed of light in the gas. On the other hand, for
a strongly bound system, such as a solid or liquid transparent
medium, the small energy flow generated by the motion with respect
to the vacuum condensate should dissipate mainly by heat conduction
with no particle flow and no light anisotropy in the rest frame of
the medium, in agreement with the classical experiments in glass and
water.

For this reason, one should design a new class of ether-drift
experiments where two optical cavities are filled with a gas and
study the frequency shift $\Delta \nu$ between the two resonators
that gives a measure of the possible anisotropy of the two-way speed
of light. Such a type of "non-vacuum" experiment would be along the
lines of Ref.\cite{holger} where just the use of optical cavities
filled with different materials was considered as a useful tool to
study possible deviations from Lorentz invariance.

The aim of this paper is to give a set of precise predictions for
this new class of ether-drift experiments. In Sect.2 we shall
provide a definite model for the two-way speed of light. In Sect.3,
we shall discuss various experimental set up and the expected form
of the signal. Finally, in Sect.4 we shall present our summary and
conclusions.

\section{The two-way speed of
light in a gaseous medium}

Rigorous treatments of light propagation in dielectric media are
based on the extinction theory \cite{born}. This was originally
formulated for continuous media where the interparticle distance is
smaller than the light wavelength. In the opposite case of an
isotropic, dilute random medium \cite{weber}, it is relatively easy
to compute the scattered wave in the forward direction and obtain
the refractive index. However, if there are convective currents,
taking into account the motion of the molecules that make up the gas
is a non-trivial problem. If solved, one expects an angular
dependence of the refractive index and an anisotropy of the phase
speed of the refracted light.

This expectation derives from a much simpler, semi-quantitative
approach where one introduces from scratch the refractive index
${\cal N}$ of the gas and the time $t$ spent by refracted light to
cover some given distance $L$ within the medium. By assuming
isotropy, one would find $t={\cal N}L/c$. This can be expressed as
the sum of $t_0=L/c$ and $t_1=({\cal N}-1)L/c$ where $t_0$ is the
same time as in the vacuum and $t_1$ represents the additional,
average time by which the refracted light is ``slowed'' down by the
presence of matter. If there are convective currents, so that $t_1$
is different in different directions, one can deduce an anisotropy
of the speed of light proportional to $({\cal N}-1)$. To see this,
let us consider light propagating in a 2-dimensional plane and
express $t_1$ as
\begin{equation} t_1={{L}\over{c}}f({\cal N}, \theta, \beta)
\end{equation} with $\beta=V/c$, $V$ being the velocity of the
laboratory with respect to the preferred frame $\Sigma$ where the
isotropic form
\begin{equation}
\label{boundary} f({\cal N}, \theta, 0)={\cal N}-1
\end{equation}
is assumed. By expanding around ${\cal N}=1$ where, whatever
$\beta$, $f$ vanishes by definition, one finds for gaseous systems
(where ${\cal N}-1 \ll 1$) the universal trend
\begin{equation} f({\cal N}, \theta,\beta)\sim ({\cal N}-1)F(\theta,\beta) \end{equation}
with
\begin{equation}
F(\theta,\beta)\equiv (\partial f/\partial {\cal N})|_{ {\cal N}=1}
\end{equation} and $F(\theta,0)=1$.
Therefore, from \begin{equation} t({\cal N},\theta,\beta)=
{{L}\over{c({\cal N},\theta,\beta)}}\sim {{L}\over{c}}+
{{L}\over{c}}({\cal N}-1)~F(\theta,\beta)
\end{equation} one gets
\begin{equation}
c({\cal N},\theta,\beta)\sim {{c}\over{ {\cal N} }}~  \left[1-
({\cal N}-1) ~(F(\theta,\beta) -1)\right]
\end{equation}
Analogous relations hold for the two-way speed of light $
\bar{c}({\cal N},\theta,\beta)$  \begin{equation} \bar{c}({\cal
N},\theta,\beta)={{2~c({\cal N},\theta,\beta)c({\cal N},\pi
+\theta,\beta)}\over{c({\cal N},\theta,\beta) +c ({\cal N},\pi
+\theta,\beta)}} \sim {{c}\over{ {\cal N} }} \left[1-  ({\cal N}-1)
 ~\left( {{F(\theta,\beta) + F(\pi+\theta,\beta)}\over{2}} -1\right) \right]
\end{equation} that is commonly measured in optical resonators. In
this case, one predicts a non-zero anisotropy
\begin{equation} {{\Delta \bar{c}_\theta}\over{c}} \equiv {{{\bar{c}}({\cal N},\pi/2,\beta)-
{\bar{c}}({\cal N},0,\beta)}\over{c}}\sim ({\cal N}-1)~{{\Delta
F}\over{2}}
\end{equation} with $\Delta F= F(0,\beta)+F(\pi,\beta)
-F(\pi/2,\beta)-F(3\pi/2,\beta)$ and the characteristic scaling law
\begin{equation} \label{scale} {{
\Delta\bar{c}_\theta( {\cal N} ) } \over{ \Delta \bar{c}_\theta(
{\cal N}') }} \sim {{ {\cal N}-1 }\over{ {\cal N}'-1 }}
\end{equation} More quantitative estimates can be obtained by exploring
some general properties of the function
$F(\theta,\beta)$. By expanding in powers of $\beta$
\begin{equation}
F(\theta,\beta)-1 = \beta F_1(\theta) + \beta^2 F_2(\theta)+...
\end{equation}
and taking into account that, by the very definition of two-way
speed, $\bar{c}({\cal N},\theta,\beta)= \bar{c}({\cal
N},\theta,-\beta)$, it follows that $F_1(\theta)=-F_1(\pi +
\theta)$. Therefore, we get the general structure of the two-way
speed of light to ${\cal O}(\beta^2)$
\begin{equation}
\label{legendre} \bar{c}({\cal N},\theta,\beta) \sim {{c}\over{
{\cal N} }} \left[1- ({\cal N}-1)~\beta^2
\sum^\infty_{n=1}\zeta_{2n}P_{2n}(\cos\theta)
  \right]
\end{equation}
in which we have expressed the combination $F_2(\theta) + F_2(\pi
+\theta)$ as an infinite expansion of even-order Legendre
polynomials with unknown coefficients $\zeta_{2n}={\cal O}(1)$.

This general structure can be compared with the corresponding result
\cite{pla} obtained by using Lorentz transformations to connect S'
to the preferred frame
\begin{equation} \label{twoway} \bar{c}({\cal N},\theta,\beta)\sim {{c}\over{ {\cal N}
}}~[1-\beta^2~(A+B\sin^2\theta)] \end{equation} with
\begin{equation} \label{lorentz} A\sim 2({\cal N}-1)
~~~~~~~~~~~~B\sim -3({\cal N}-1)\end{equation} that corresponds to
set in Eq.(\ref{legendre}) $\zeta_2=2$ and all $\zeta_{2n}=0$ for $n
> 1$. Eqs.(\ref{twoway})-(\ref{lorentz}), that represent a definite realization
of the general structure in (\ref{legendre}), provide a partial
answer to the problems posed by our limited knowledge of the
electromagnetic properties of gaseous systems and will be adopted in
the following as our basic model for the two-way speed of light.

Notice that Eqs.(\ref{twoway})-(\ref{lorentz}) lead to
\begin{equation} \label{eq1}
{{\Delta\bar{c}_\theta( {\cal N})}\over{c}} \sim 3 ({\cal N}
-1)~{{V^2}\over{c^2}}\end{equation} and thus Eq.(\ref{scale}) is
identically satisfied. At the same time, one gets agreement with the
pattern observed in the classical and modern ether-drift
experiments, as illustrated in Refs.\cite{pla}, that suggests (for
gaseous media {\it only}) a relation of the type in Eq.(\ref{eq1}).
In fact, in the classical experiments performed in air at
atmospheric pressure, where ${\cal N}\sim 1.000293$, the observed
anisotropy was ${{\Delta\bar{c}_\theta}\over{c}}\lesssim 10^{-9}$
thus providing a typical value $V/c\sim 10^{-3}$, as that associated
with most cosmic motions. Analogously, in the classical experiments
performed in helium at atmospheric pressure, where ${\cal N}\sim
1.000035$ (and in a modern experiment with He-Ne lasers where ${\cal
N}\sim 1.00004$), the observed effect was
${{\Delta\bar{c}_\theta}\over{c}}\lesssim 10^{-10}$ so that again
$V/c\sim 10^{-3}$.

Notice also that, although originating from a different theoretical
framework, Eq.(\ref{twoway}) is formally analogous to the expression
of the two-way speed of light in the RMS formalism
\cite{robertson,ms} where $A$ and $B$ are taken as free parameters.

One conceptual detail concerns the gas refractive index whose
reported values are experimentally measured on the earth by two-way
measurements. For instance for the air, the most precise
determinations are at the level $10^{-7}$, say ${\cal N}_{\rm
air}=1.0002926..$ for yellow light at STP (Standard Temperature and
Pressure). By assuming a non-zero anisotropy in the earth's frame,
one should interpret the isotropical value $c/{\cal N_{\rm air}}$ as
an angular average of Eq.(\ref{twoway}), i.e.
\begin{equation} \label{nair} {{c}\over{ {\cal N}_{\rm air} }}\equiv
\langle\bar{c}(\bar{\cal N}_{\rm air},\theta,\beta)\rangle=
{{c}\over{ \bar {\cal N} _{\rm air} }} ~[1-{{1}\over{2}} (\bar{\cal
N}_{\rm air} -1){{V^2}\over{c^2}}]
\end{equation} From this relation, one can determine the unknown value $\bar
{\cal N} _{\rm air} \equiv {\cal N}(\Sigma)$ (as if the gas were at
rest in $\Sigma$), in terms of the experimentally known quantity
${\cal N}_{\rm air}\equiv{\cal N}(earth)$ and of $V$. In practice,
for the standard velocity values involved in most cosmic motions,
say 200 km/s $\leq V \leq $ 400 km/s, the difference between ${\cal
N}(\Sigma)$ and ${\cal N}(earth)$ is well below $10^{-9}$ and thus
completely negligible. The same holds true for the other gaseous
systems at STP (say nitrogen, carbon dioxide, helium,..) for which
the present experimental accuracy in the refractive index is, at
best, at the level $10^{-6}$. Finally, the isotropic two-way speed
of light is better determined in the low-pressure limit where
$({\cal N}-1)\to 0$. In the same limit, for any given value of $V$,
the approximation ${\cal N}(\Sigma)={\cal N}(earth)$ becomes better
and better.

\section{Ether-drift experiments in gaseous media}

From the point of view of ether-drift experiments, the crucial
ingredient, that might indicate the existence of a preferred frame,
consists in detecting the characteristic modulations of the signal
due to the earth's rotation. Descriptions of this important effect
are already available in the literature. For instance, within the
SME model \cite{sme} the relevant formulas are given in the appendix
of Ref.\cite{mewes} and for the RMS test theory \cite{robertson,ms}
one can look at Ref.\cite{applied}. However, either due to the great
number of free parameters (19 in the SME model) and/or to the
restriction to a definite experimental set up, it is not always easy
to adapt these papers to the actual conditions needed for our
experimental test. For this reason, in the following, we will
present a set of compact formulas that can be immediately used by
the reader to evaluate the signal when two arbitrary gaseous media
fill the resonating cavities. The formalism covers most experimental
set up including the very recent type of experiment proposed in
Ref.\cite{luiten} to perform tests of the Standard Model.

The main point is that the earth's rotation enters only through two
quantities, $v=v(t)$ and $\theta_0=\theta_0(t)$, respectively the
magnitude and the angle associated with the projection of the
unknown cosmic earth's velocity ${\bf{V}}$ in the plane of the
interferometer.

Once the angle $\theta_0$ is conventionally defined when one of the
arms of the interferometer is oriented to the North point in the
laboratory (counting $\theta_0$ from North through East so that
North is $\theta_0=0$ and East is $\theta_0=\pi/2$), we can
immediately use the formulas given by Nassau and Morse
\cite{nassau}. These are valid for short-term observations, say 3-4
days, where there are no appreciable changes in the cosmic velocity
due to changes in the earth's orbital velocity around the Sun and
the only time dependence is due to the earth's rotation.

In this approximation, introducing the magnitude $V$ of the full
earth's velocity with respect to a hypothetic preferred frame
$\Sigma$, its right ascension $\alpha$ and angular declination
$\gamma$, we get
\begin{equation} \label{cosine}
       \cos z(t)= \sin\gamma\sin \phi + \cos\gamma
       \cos\phi \cos(\tau-\alpha)
\end{equation} \begin{equation}
       \sin z(t)\cos\theta_0(t)= \sin\gamma\cos \phi -\cos\gamma
       \sin\phi \cos(\tau-\alpha)
\end{equation} \begin{equation}
       \sin z(t)\sin\theta_0(t)= \cos\gamma\sin(\tau-\alpha) \end{equation}
\begin{equation} \label{projection}
       v(t)=V \sin z(t) ,
\end{equation}
Here $z=z(t)$ is the zenithal distance of ${\bf{V}}$. Namely, $z=0$
corresponds to a ${\bf{V}}$ which is perpendicular to the plane of
the interferometer and $z=\pi/2$ to a ${\bf{V}}$ that lies entirely
in that plane. Further, $\phi$ is the latitude of the laboratory and
$\tau=\omega_{\rm sid}t$ is the sidereal time of the observation in
degrees ($\omega_{\rm sid}\sim {{2\pi}\over{23^{h}56'}}$).

Let us now consider two orthogonal cavities oriented for simplicity
North-South (cavity 1) and East-West (cavity 2) in the laboratory
frame. They are filled with two different gaseous media with
refractive indices ${\cal N}_i$ (i=1,2) such that ${\cal
N}_i=1+\epsilon_i$, and $0\leq \epsilon_i \ll 1$. The frequency in
each cavity is \begin{equation} \nu_i(\theta_i)=\bar{c}_i({\cal
N}_i,\theta_i,\beta)k_i \end{equation} and the frequency shift is
\begin{equation} \Delta\nu=\nu_1(\theta_1)-\nu_2(\theta_2)
\end{equation} In the above relations we have introduced the
parameters $k_i$
\begin{equation} k_i={{m_i}\over{2L_i}}\end{equation} where $m_i$ are integers
fixing the cavity modes and $L_i$ are the cavity lengths. Finally,
 $\theta_i$ is the angle
between ${\bf{V}}$ and the axis of the i-th cavity and
$\bar{c}_i({\cal N}_i,\theta_i,\beta)$ denote the two-way speeds of
light in (\ref{twoway}).

We observe that, in the presence of an effective vacuum thermal
gradient, one might also consider pure thermal conduction effects in
the solid parts of the apparatus. Even by using cavities with an
ultra-low thermal expansion coefficient, these conduction effects
could induce tiny differences of the cavity lengths (and thus of the
cavity frequencies) upon active rotations of the apparatus or under
the earth's rotation. However, this effect does not depend on the
gas that fills the cavity and therefore can be preliminarily
evaluated and subtracted out by first running the experiment in the
vacuum mode, i.e. at the same room temperature but when no gas is
present inside the cavities. The precise experimental limits of
Ref.\cite{herrmann} (obtained with vacuum cavities at room
temperature) show that any such effect can be reduced to the level
$10^{-15}-10^{-16}$ and thus would be irrelevant for our purpose. In
fact, as we shall show in a moment, the typical magnitudes of the
signal, expected by running the experiments in the gaseous mode,
should be larger by 4-5 orders of magnitude.

By introducing the unit vectors $\hat{\bf n}_i$ that fix the
direction of the two cavities and the projection ${\bf{v}}$ of the
full ${\bf{V}}$ in the interferometer's plane, one finds
\begin{equation}
V^2\sin^2\theta_i=V^2(1-\cos^2\theta_i)=V^2-(\hat{\bf n}_i\cdot
{\bf{v}} )^2 \end{equation} so that ($v=|{\bf{v}}|$)
\begin{equation} V^2\sin^2\theta_1=V^2- v^2\cos^2\theta_0 \end{equation} and
\begin{equation} V^2\sin^2\theta_2=V^2- v^2\sin^2\theta_0\end{equation}
Therefore, by defining the reference frequency $\nu_0={{c
k_1}\over{{\cal N}_1}}$ and introducing the parameter $\xi$ through
\begin{equation} \xi={{ {\cal N}_1 k_2}\over{{\cal N}_2 k_1 }} \end{equation}
one finds the relative frequency shift \begin{equation}
\label{general} {{\Delta \nu(t)}\over{\nu_0}}=1- \xi +
{{V^2}\over{c^2}}[\xi(A_2 +B_2) -(A_1+B_1)] +
{{v^2(t)}\over{c^2}}[B_1\cos^2\theta_0(t) - \xi
B_2\sin^2\theta_0(t)] \end{equation}  For a symmetric apparatus
where ${\cal N}_1={\cal N}_2$, $A_1=A_2$, $B_1=B_2=B$ and $\xi=1$,
one finds
\begin{equation} \label{symm}{{\Delta \nu(t)_{\rm
symm}}\over{\nu_0}} = B {{v^2(t)}\over{c^2}} \cos2\theta_0(t)
\end{equation} On the other hand for a non-symmetric apparatus of
the type considered in Ref.\cite{luiten} with $L_1=L_2=L$, but where
one can conveniently arrange ${\cal N}_1=1$ (up to negligible terms)
so that $A_1\sim B_1 \sim 0$, denoting ${\cal N}_2={\cal N}$,
$A_2=A$, $B_2=B$, ${{m_2}\over{m_1}}={\cal P}$, we find
\begin{equation} \label{asymm} {{\Delta \nu(t)}\over{\nu_0}}=1-
{{{\cal P}\over{\cal N}}}+ {{{\cal P}\over{\cal
N}}}{{V^2}\over{c^2}}(A +B) - B{{{\cal P}\over{\cal
N}}}{{v^2(t)}\over{c^2}} \sin^2\theta_0(t) \end{equation} To
consider experiments where one or both resonators are placed in a
state of active rotation (at a frequency $\omega_{\rm rot} \gg
\omega_{\rm sid}$), it is convenient to modify Eq.(\ref{general}) by
rotating the resonator 1 by an angle $\delta_1$ and the resonator 2
by an angle $\delta_2$ so that the last term in Eq.(\ref{general})
becomes
\begin{equation} {{v^2(t)}\over{
c^2}}[B_1\cos^2(\delta_1-\theta_0(t)) - \xi
B_2\sin^2(\delta_2-\theta_0(t))] \end{equation} Therefore, in a
fully symmetric apparatus where ${\cal N}_1={\cal N}_2$, $A_1=A_2$,
$B_1=B_2=B$ and $\xi=1$ and both resonators rotate, as in
Ref.\cite{schiller}, setting
\begin{equation}\delta_1=\delta_2=\omega_{\rm rot}t \end{equation} one obtains
\begin{equation} \label{symm2}  {{\Delta \nu(t)_{\rm
symm}}\over{\nu_0}}= B {{v^2(t)}\over{c^2}} \cos2( \omega_{\rm rot}t
-\theta_0(t)) \end{equation} On the other hand, if only one
resonator rotates, as in Ref.\cite{herrmann}, setting $\delta_1=0$
and $\delta_2=\omega_{\rm rot}t$ one obtains the alternative result
\begin{equation} \label{asymm2} {{\Delta \nu(t)}\over{\nu_0}}= B
{{v^2(t)}\over{2 c^2}}[\cos2\theta_0(t) + \cos2( \omega_{\rm rot}t
-\theta_0(t))] \end{equation} By first filtering the signal at the
frequency $\omega=\omega_{\rm rot} \gg \omega_{\rm sid}$, the main
difference between the two expressions is an overall factor of two.

Let us now return to the general case of a non-rotating set up
Eq.(\ref{general}). Using Eqs.({\ref{cosine})-({\ref{projection}) we
obtain the simple Fourier expansion \begin{equation} {{\Delta
\nu(t)}\over{\nu_0}}=1-\xi + (g_0+g_1\sin\tau +g_2\cos\tau +g_3\sin
2\tau+ g_4\cos2\tau )\end{equation} where \begin{equation}
\label{g0} g_0={{V^2}\over{c^2}}[ \xi(A_2 +B_2) -(A_1+B_1) +
B_1(\sin^2\gamma\cos^2\phi+ {{1}\over{2}}\cos^2\gamma\sin^2\phi)
-{{1}\over{2}}\xi B_2 \cos^2\gamma ]\end{equation} \begin{equation}
\label{f12} g_1=-{{1}\over{2}}{{V^2}\over{c^2}}B_1\sin 2\gamma\sin
2\phi \sin \alpha
~~~~~~~~~~~~~~~~~~~~~~~g_2=-{{1}\over{2}}{{V^2}\over{c^2}}B_1\sin
2\gamma\sin 2\phi \cos \alpha \end{equation} \begin{equation}
\label{f34} g_3={{1}\over{2}}{{V^2}\over{c^2}}(B_1\sin^2\phi +\xi
B_2)\cos^2\gamma\sin 2\alpha
~~~~~~~~~~g_4={{1}\over{2}}{{V^2}\over{c^2}}(B_1\sin^2\phi +\xi
B_2)\cos^2\gamma\cos 2\alpha~~\end{equation} Since the mean signal
is most likely affected by systematic effects, one usually
concentrates on the daily modulation. In this case, assuming that
$g_1$, $g_2$, $g_3$ and $g_4$ can be extracted to good accuracy from
the experimental data, one can try to obtain a pair of angular
variables through the two independent determinations of $\alpha$
\begin{equation} \label{alpha} \tan \alpha=
{{g_1}\over{g_2}}~~~~~~~~~~~~~~~\tan 2\alpha=
{{g_3}\over{g_4}}\end{equation} and the relation \begin{equation}
\tan |\gamma| ={{|B_1\sin^2\phi +\xi B_2|}\over{|2 B_1\sin 2\phi|}}~
\sqrt{ {{g^2_1+g^2_2}\over{g^2_3+g^2_4}}} \end{equation} Notice that
Eqs.(\ref{g0})-(\ref{f34}) remain unchanged under the replacement
$(\alpha,\gamma) \to (\alpha+\pi,-\gamma)$. Also, two dynamical
models that predict the same anisotropy parameters up to an overall
re-scaling $B_i \to \lambda B_i$ would produce the same $|\gamma|$
from the experimental data.

Finally for a symmetric apparatus, where $B_1=B_2=B$ and $\xi=1$,
one obtains the simpler relation \begin{equation} \label{gamma} \tan
|\gamma| ={{1+\sin^2\phi }\over{|2 \sin 2\phi|}}~ \sqrt{
{{g^2_1+g^2_2}\over{g^2_3+g^2_4}}} \end{equation} where any
reference to the anisotropy parameters drops out.

To obtain some order of magnitude estimate, let us consider the
amplitude of the modulation of the signal at the sidereal frequency
for a typical latitude of the laboratory $|\phi |\sim 45^o$. This is
given by \begin{equation} g_{\omega_{\rm sid}}= \sqrt{ g^2_1 +
g^2_2}= {{1}\over{2}}{{V^2}\over{c^2}}~|B_1\sin 2\gamma|
\end{equation} By assuming the cavity 1 to be filled with carbon
dioxide (whose refractive index at atmospheric pressure is ${\cal
N}_1 \sim 1.00045$) and the typical value ${{V^2}\over{c^2}} \sim
10^{-6}$ (associated with most cosmic motions) one expects a typical
modulation of the relative frequency shift $ g_{\omega_{\rm
sid}}\sim 10^{-10}$. Analogously, for helium at atmospheric pressure
(where ${\cal N}_1 \sim 1.000035$) one expects $g_{\omega_{\rm
sid}}\sim 10^{-11}$. As anticipated, these values would be 4$-$5
orders of magnitude larger than the limit $10^{-15}-10^{-16}$ placed
by the present ether-drift experiments in vacuum.

\section{Summary and conclusions}

In principle, on the basis of very general arguments related to a
non-zero vacuum energy, the physical condensed vacuum of present
particle physics might represent a preferred reference frame. In
this case, in any moving frame there might be a non-zero vacuum
energy-momentum flow along the direction of motion. By treating the
quantum vacuum as a relativistic medium, this non-zero
energy-momentum flow should behave as an effective thermal gradient.
As such, it could induce small convective currents in a loosely
bound system as a gas and an anisotropy of the speed of light.

For this reason, we have considered in this paper a new class of
ether-drift experiments in which optical resonators are filled by
gaseous media. The existence of convective currents leads to the
general structure of the two-way speed in Eq.(\ref{legendre}) that
admits Eqs.(\ref{twoway})-(\ref{lorentz}) as a special case.

In this particular limit, by using the basic relations
(\ref{cosine})-(\ref{projection}) to take into account the effect of
the earth's rotation, we have derived a set of definite predictions
that cover most experimental set up. For the typical velocities
involved in most cosmic motions, the expected relative frequency
shift between the two resonators should be about 4$-$5 orders of
magnitude larger than the limit $10^{-15}-10^{-16}$ placed by the
present ether-drift experiments in vacuum.

We want to emphasize that, due to the limited precision
characterizing our knowledge of the electromagnetic properties of
gaseous media, that forces us to restrict to relations
(\ref{twoway})-(\ref{lorentz}) for the two-way speed, we cannot
exclude the existence of other competing mechanisms that, while
physically different from our proposed drift of the vacuum energy,
may simulate the same effects. For instance, a similar direction
dependence of the refractive index might also be introduced if the
molecules in the gas exhibit no net motion but instead a suitable
non-isotropic local interaction of the incoming radiation with the
medium is introduced, for instance within the more general framework
of the SME model \cite{sme}. In this case, there might be
non-equivalent ways to obtain the same characteristic experimental
signatures.

Still, we believe that our picture of light anisotropy, as arising
from the convective currents that can be established in dilute
systems, provides a simple theoretical framework to understand why
Eq.(\ref{eq1}), while being consistent with the pattern observed in
gaseous systems, does {\it not} apply to Michelson-Morley
experiments  performed in solid transparent media \cite{fox} as
perspex (where ${\cal N}\sim 1.5$).

In any case, exploring the class of scenarios consistent with
Eqs.(\ref{twoway})-(\ref{lorentz}) leads to consider the following
experimental checks:

~~i) for a symmetric apparatus one should try to extract from the
data the product $H=B{{V^2}\over{c^2}}$ and, by using
Eqs.(\ref{alpha}) and (\ref{gamma}), two pairs of conjugate angular
variables $(\alpha,\gamma)$ $(\alpha+\pi,-\gamma)$. Also, by
suitably changing the gaseous medium (and its pressure) within the
cavities, one should try to check the precise trend predicted in
Eqs.(\ref{scale}) and (\ref{lorentz}), namely
\begin{equation} {{H'}\over{H''}}\sim {{ {\cal N}'-1}\over{ {\cal
N}''-1 }}\end{equation}

ii)~~for a non-symmetric apparatus of the type proposed in
Ref.\cite{luiten}, where one can conveniently fix the cavity
oriented North-South to have ${\cal N}_1=1$ (up to negligible
terms), by using Eqs.(\ref{lorentz}) one predicts $B_1\sim 0$ in
Eqs.(\ref{f12}) and (\ref{f34}) so that all time dependence should
be due to $B_2$. Thus the modulation of the signal should be a pure
$\omega=2\omega_{\rm sid}$ effect with no appreciable contribution
at $\omega=\omega_{\rm sid}$

iii)~~for a deeper analysis, one should keep in mind that, in each
single session, the direction $(\alpha,\gamma)$ cannot be
distinguished from the opposite direction $(\alpha+\pi,-\gamma)$.
For this reason, a whole set j=1,2..M of short-term experimental
sessions should be performed in different periods along the earth's
orbit to obtain an overall consistency check. Notice that, for a
complete description of the observations over a one-year period, it
is not necessary to modify the simple formulas
Eqs.(\ref{g0})-(\ref{f34}) and introduce explicitly the further
modulations associated with the orbital frequency $\Omega_{\rm
orb}\sim {{2\pi}\over{1 ~{\rm year}}}$. Rather, by plotting on the
celestial sphere all directions defined by the $(\alpha_j,\gamma_j)$
pairs obtained in the various short-term observations, one can try
to reconstruct the earth's ``aberration circle''. If this will show
up, by using the formulas of the spherical triangles, one will be
able to determine the mean cosmic velocity $\langle V\rangle $ from
the angular opening of the circle and the known value of the earth's
orbital
 velocity $\sim 30 $ km/s. In this way, given the value of
$\langle H \rangle$, one will be able to disentangle $\langle
V\rangle $ from $B$ and estimate the absolute magnitude of the
anisotropy parameter.


\vskip 50 pt
\begin{thebibliography} {99}
 \bibitem{thooft}
G. 't Hooft, In Search of the Ultimate Building Blocks, Cambridge
Univ. Press, Cambridge 1997, page 70.
\bibitem{mech}
M. Consoli and P.M. Stevenson, Int. J. Mod. Phys. {\bf A15}, 133
(2000), hep-ph/9905427.
\bibitem{pagano}
M. Consoli, A. Pagano and L. Pappalardo, Phys. Lett. {\bf A318}, 292
(2003).
\bibitem{epjc}
M. Consoli and E. Costanzo, Eur. Phys. J. {\bf C54}, 285 (2008).
\bibitem{cpt}
See, for instance, R. F. Streater and  A. S. Wightman, PCT, Spin and
Statistics, and all that, W. A. Benjamin, New York 1964.
\bibitem{zeldovich}
Y. B. Zeldovich, Sov. Phys. Usp. {\bf 11}, 381 (1968).
\bibitem{weinberg}
S. Weinberg, Rev. Mod. Phys. {\bf 61}, 1 (1989).
\bibitem{ciancitto}
M. Consoli and A. Ciancitto, Nucl. Phys. {\bf B254}, 653 (1985).
\bibitem{poincare1}
E. V. Stefanovich, Found. Phys. {\bf 32}, 673 (2002).
\bibitem{poincare2}
S. D. Glazek and T. Maslowksi, Phys. Rev. {\bf 65}, 065011 (2002).
\bibitem{volovik}
G. E. Volovik, Phys. Rep. {\bf 351}, 195 (2001).
\bibitem{adler}
For a general review, see S.~L.~Adler, Rev. Mod. Phys. {\bf 54}, 729
(1982).
\bibitem{holger}
H. M\"uller, Phys. Rev. {\bf D71}, 045004 (2005).
\bibitem{born}
M. Born and E. Wolf, Principles of Optics, Cambridge University
Press, Cambridge 1999, pp. 103-115.
\bibitem{weber}
V. C. Ballenegger and T. A. Weber, Am. J. Phys.{\bf 67}, 599 (1999).
\bibitem{pla}
M. Consoli and E. Costanzo, Phys. Lett. {\bf A333}, 355 (2004); N.
Cim. {\bf 119B}, 393 (2004) [arXiv:gr-qc/0406065].
\bibitem{robertson}
H. P. Robertson, Rev. Mod. Phys. {\bf 21}, 378 (1949).
\bibitem{ms}
R. M. Mansouri and R. U. Sexl, Gen. Rel. Grav. {\bf 8}, 497 (1977).
\bibitem{sme}
D. Colladay and V. A. Kostelecky, Phys. Rev. {\bf D55}, 6760 (1997);
{\bf D58}, 111602 (1998); R. Bluhm, et al., Phys. Rev. Lett. {\bf
88}, 090801 (2002).
\bibitem{mewes}
V. A. Kostelecky and M. Mewes, Phys. Rev. {\bf D66}, 056005 (2002).
\bibitem{applied}
H. M\"uller, et al. Appl. Phys. {\bf B77}, 719 (2003).
\bibitem{luiten}
S. Dawkins and A. Luiten, "Testing the standard model of physics",
Presentation at the Australian Institute of Physics 17th National
Congress 2006, Brisbane December 2006.
\bibitem{nassau}
J. J. Nassau and P. M. Morse, Astrophys. Journ. {\bf 65}, 73 (1927).
\bibitem{herrmann}
S. Herrmann, et al., Phys. Rev. Lett. {\bf 95}, 150401 (2005)
[arXiv:physics/0508097].
\bibitem{schiller}
P. Antonini, et al., Phys. Rev. {\bf A71}, 050101(R) (2005)
[arXiv:gr-qc/0504109].
\bibitem{fox}
J. Shamir and R. Fox, N. Cim. {\bf 62B}, 258 (1969).









\end{thebibliography}
\end{document}